\documentclass[review]{elsarticle}

\usepackage{lineno,hyperref}
\usepackage{amsmath}
\usepackage{xcolor}
\modulolinenumbers[5]

\journal{Astroparticle Physics}

\begin{document}

\begin{frontmatter}

\title{Lowering the energy threshold in COSINE-100 dark matter searches}

\author{The COSINE-100 Collaboration\hspace{15em}}
\address{}


\author[a,p]{G.~Adhikari}
\author[b]{E.~Barbosa~de~Souza}
\author[c]{N.~Carlin}
\author[d]{J.~J.~Choi}
\author[d]{S.~Choi}
\author[e]{M.~Djamal}
\author[f]{A.~C.~Ezeribe}
\author[c]{L.~E.~Fran{\c c}a}
\author[g]{C.~Ha\corref{corr1}}
\cortext[corr1]{Corresponding author}
\ead{chha@cau.ac.kr}
\author[h]{I.~S.~Hahn}
\author[i]{E.~J.~Jeon}
\author[b]{J.~H.~Jo}
\author[i]{W.~G.~Kang}
\author[j]{M.~Kauer}
\author[i]{H.~Kim}
\author[k]{H.~J.~Kim}
\author[i]{K.~W.~Kim}
\author[d]{S.~K.~Kim}
\author[i,a,l]{Y.~D.~Kim}
\author[i,m,l]{Y.~H.~Kim}
\author[i]{Y.~J.~Ko\corref{corr2}}
\cortext[corr2]{Corresponding author}
\ead{yjko@ibs.re.kr}
\author[i]{E.~K.~Lee}
\author[i,l]{H.~S.~Lee}
\author[i]{J.~Lee}
\author[k]{J.~Y.~Lee}
\author[i,l]{M.~H.~Lee}
\author[l,i]{S.~H.~Lee}
\author[i]{D.~S.~Leonard}
\author[c]{B.~B.~Manzato}
\author[b]{R.~H.~Maruyama}
\author[f]{R.~J.~Neal}
\author[i]{S.~L.~Olsen}
\author[l,i]{B.~J.~Park}
\author[n]{H.~K.~Park}
\author[m]{H.~S.~Park}
\author[i]{K.~S.~Park}
\author[c]{R.~L.~C.~Pitta}
\author[i]{H.~Prihtiadi}
\author[i]{S.~J.~Ra}
\author[o]{C.~Rott}
\author[i]{K.~A.~Shin}
\author[f]{A.~Scarff}
\author[f]{N.~J.~C.~Spooner}
\author[b]{W.~G.~Thompson}
\author[p]{L.~Yang}
\author[o]{G.~H.~Yu}

\address[a]{Department of Physics, Sejong University, Seoul 05006, Republic of Korea}
\address[b]{Department of Physics and Wright Laboratory, Yale University, New Haven, CT 06520, USA}
\address[c]{Physics Institute, University of S\~{a}o Paulo, 05508-090, S\~{a}o Paulo, Brazil}
\address[d]{Department of Physics and Astronomy, Seoul National University, Seoul 08826, Republic of Korea}
\address[e]{Department of Physics, Bandung Institute of Technology, Bandung 40132, Indonesia}
\address[f]{Department of Physics and Astronomy, University of Sheffield, Sheffield S3 7RH, United Kingdom}
\address[g]{Department of Physics, Chung-Ang University, Seoul 06973, Republic of Korea}
\address[h]{Department of Science Education, Ewha Womans University, Seoul 03760, Republic of Korea} 
\address[i]{Center for Underground Physics, Institute for Basic Science (IBS), Daejeon 34126, Republic of Korea}
\address[j]{Department of Physics and Wisconsin IceCube Particle Astrophysics Center, University of Wisconsin-Madison, Madison, WI 53706, USA}
\address[k]{Department of Physics, Kyungpook National University, Daegu 41566, Republic of Korea}
\address[l]{IBS School, University of Science and Technology (UST), Daejeon 34113, Republic of Korea}
\address[m]{Korea Research Institute of Standards and Science, Daejeon 34113, Republic of Korea}
\address[n]{Department of Accelerator Science, Korea University, Sejong 30019, Republic of Korea}
\address[o]{Department of Physics, Sungkyunkwan University, Suwon 16419, Republic of Korea}
\address[p]{Department of Physics, University of California San Diego, La Jolla, CA 92093, USA}

\begin{abstract}
COSINE-100 is a dark matter detection experiment that uses NaI(Tl) crystal detectors operating at the Yangyang underground laboratory in Korea since September 2016. Its main goal is to test the annual modulation observed by the DAMA/LIBRA experiment with the same target medium. Recently DAMA/LIBRA has released data with an energy threshold lowered to 1 keV, and the persistent annual modulation behavior is still observed at 9.5$\sigma$. By lowering the energy threshold for electron recoils to 1~keV, COSINE-100 annual modulation results can be compared to those of DAMA/LIBRA in a model-independent way. Additionally, the event selection methods provide an access to a few to sub-GeV dark matter particles using constant rate studies. In this article, we discuss the COSINE-100 event selection algorithm, its validation, and efficiencies near the threshold.
\end{abstract}


\end{frontmatter}

\section{Introduction}
\label{sec:intro}
A quarter of the total mass-energy of universe is thought to be dark matter, as has been evidenced by various observations over the last few decades~\cite{Clowe:2006eq,Aghanim:2018eyx}. It has been theorized that dark matter is composed of particles that interact with Standard Model particles through weakly interacting processes. According to such theories, weakly interacting massive particles (WIMPs) were thermally produced in the early universe with an abundance roughly reproducing the relic abundance of $\Omega_{CDM}=0.25$ assuming a weak, self-interaction cross section~\cite{prl_39_165,JUNGMAN1996195,Schumann:2019eaa}.

Direct detection experiments~\cite{Aprile:2018dbl,mount2017luxzeplin,Agnese:2017njq,Adhikari:2018ljm} search for signals produced by WIMPs that interact with nuclei in the target material. To date, no experiment has been successful in finding a positive signal that can be interpreted as resulting from WIMPs, with the notable exception of the DAMA/LIBRA experiment that measures an annual modulation signal from their residual count rate in the energy range of 2 to 6~keV recorded in NaI(Tl) crystal detectors~\cite{Bernabei:2013xsa}. The implication of the DAMA/LIBRA result that this annual modulation is driven by a changing flux of dark matter particles through the Earth due to the Earth's revolution around the Sun has caused a controversy~\cite{Nygren:2011,Davis:2014cja,Ralston:2010bd} and an independent verification is essential.

Recently, the DAMA/LIBRA collaboration updated their results with more statistics and an energy threshold that is lowered from 2 to 1~keV~\cite{Bernabei:2018yyw}. The new results show that the annual modulation signal persists in the extended energy range (1 to 2~keV). Experiments to test the DAMA/LIBRA results are actively being carried out by several groups\cite{Adhikari:2019off,Amare:2019jul,sabre} using the same low-background NaI(Tl) target material and reaching the same energy threshold of 1~keV electron equivalent energy. In addition to facilitating the direct comparison with the claimed modulation signal, because the expected event rate of the WIMP-induced nuclear recoil scattering off a target nucleus follows an exponentially decreasing signature as a function of the measured energy, the lowering of the threshold significantly improves the WIMP detection sensitivity, and provides coverage of smaller cross sections and masses. Here, we present an event selection procedure that enables COSINE-100 to achieve a 1~keV threshold.

\section{The COSINE-100 experiment}
\label{sec:exp}

The COSINE-100 experiment consists of eight low-background NaI(Tl) crystal detectors surrounded by layers of shielding. The crystals are cylindrical and individually encapsulated in copper and coupled to 3-inch Hamamatsu R12669SEL PMTs on each flat end surface of the cylinder. The total mass of the eight crystals is 106 kg, of which the average light yield of six crystals is about 15 photoelectrons/keV, excluding two crystals which show low light output due to crystal-to-quartz coupling issues~\cite{Adhikari:2017esn}. These crystals are submerged in 2200~liters of liquid scintillator (LS) that tags LS-crystal coincident interactions. Events that are tagged as coincident interactions can be excluded from the signal search region because of the negligible probability of WIMPs scattering twice within the detector volume due to their minuscule cross sections\footnote{The probability of random coincidence with the LS event is 0.006\%, so there is no need to consider the inefficiency of the WIMP signal.}. Additionally, the tagged events provide a control sample of events that can be used to test or fit background models independently from the WIMP analysis, which only uses single-hit events. Outside of the LS, 3-cm thick copper and 20-cm thick lead shields provide attenuation of environmental radiation. The entire array is surrounded by 37 scintillating plastic panels providing a 4$\pi$ coverage of the detectors for identifying and vetoing cosmic-ray muons. Details about the experimental setup can be found elsewhere~\cite{Prihtiadi:2017inr,Adhikari:2018fpo,Adhikari_2018,Park:2017jvs}.

PMTs are known to generate noise pulses caused by dark current, occasional flashes, and radioactivity in their adjacent dynode circuitry~\cite{PMT}. Since at low energies the rate of these noise pulses is overwhelmingly higher than that of the desired scintillation pulses, one must reject the PMT-induced noise before performing a WIMP search. Fortunately, these noise pulses have decay forms that are distinct from those for particle-generated scintillation pulses in the crystal. We describe an event selection method that achieves a noise contamination level of less than 1\% of the selected scintillation signal rate in the 1 to 1.5~keV energy bin by rejecting almost all PMT-noise induced events.

\section{Pulse Shape Discrimination for Lowering Threshold}
\label{sec:params}

\subsection{Two parameters based on pulse shape}

\begin{figure}[t]
\centering
\includegraphics[width=0.8\textwidth]{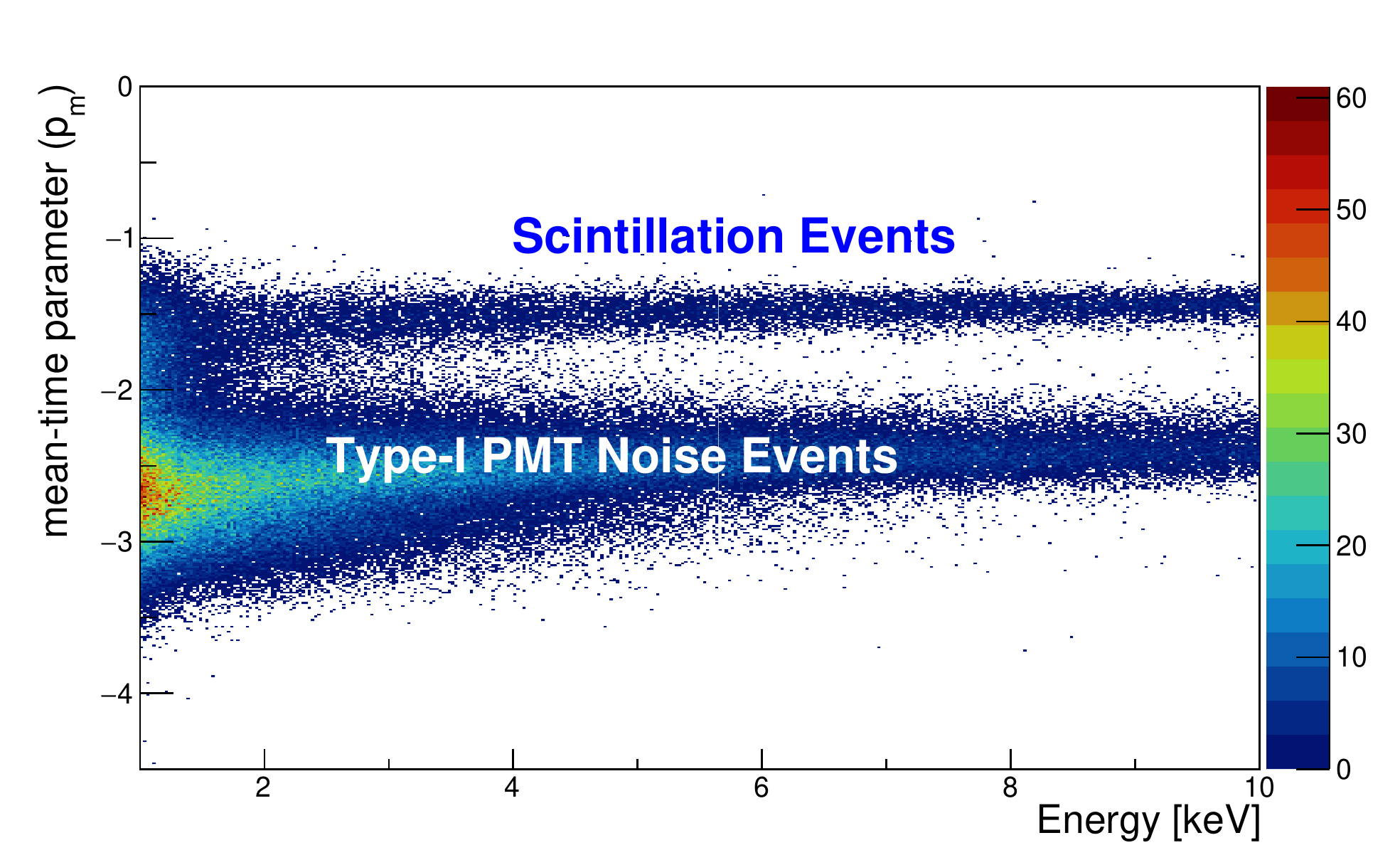}
\caption{Mean-time parameter distribution as a function of energy (59.5 days). The upper horizontal band denotes scintillation-like (scintillation) events and the lower band shows the noise-like (PMT noise) events. Below 2~keV in the high mean-time region, there are lots of noise events that cannot be separated from scintillation-like events using the mean-time parameter only.}
\label{fig:emPar}
\end{figure}

Particle-induced pulses are produced from scintillation light with the 250~ns characteristic decay time of NaI(Tl) crystals~\cite{Tanabashi:2018oca}. This decay time is longer than that for PMT noise pulses, which are typically 50~ns or less. We call this type of the PMT noise a ``type-I PMT noise'' or a thin pulse. Hence, to separate most type-I PMT-noise events from scintillation events, the amplitude-weighted mean time\footnote{$\left<t\right> \equiv \frac{\sum_it_i\cdot a_i}{\sum_ia_i}-T_1$, where $t_i$ and $a_i$ are time and amplitude (analog-to-digital counts) for $i^{th}$ bin of a pulse, respectively. $T_1$ is the time of the first photoelectron for a PMT pulse.} of the PMT pulse is calculated. For each event, the mean times of the two PMT pulses recorded in a crystal are combined into one parameter defined as $p_m \equiv \ln \left( \left< t\right>_1 + \left< t\right>_2\right)$, where $\left<t\right>_i$ is the amplitude-weighted mean time of the $i^{\mathrm{th}}$ PMT~\cite{Kim2019}.

The mean-time parameter $p_m$ provides an effective method for separating scintillation events from the type-I PMT-noise events above 2~keV (see Fig.~\ref{fig:emPar}). However, it is apparent in the figure that at energies below 2~keV, a distinct type of a new noise population occurs with mean times that extend well into the scintillation signal region. Thus, at these energies, the mean-time parameter alone does not discriminate a significant fraction of the noise which calls for additional selection criteria.

\begin{figure}[t]
\centering
\includegraphics[width=0.8\textwidth]{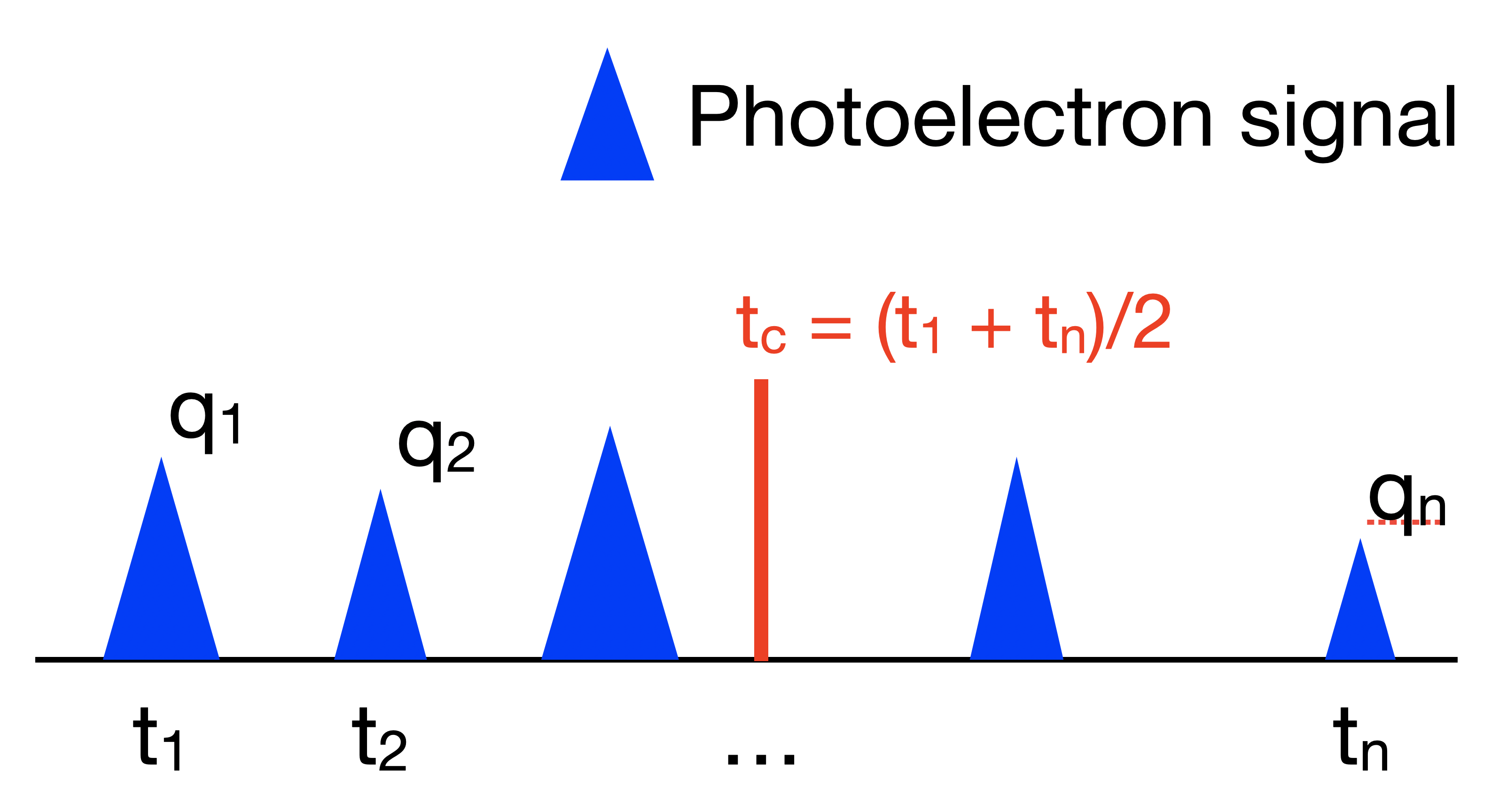}
\caption{An illustration of a PMT pulse that has several single photoelectron signals to explain the construction of the parameter defined as Eq.\ref{eq:SimplifiedDecayTime}}
\label{fig:pulse_shape_parameter}
\end{figure}

\begin{figure}[t]
\centering
\includegraphics[width=0.9\textwidth]{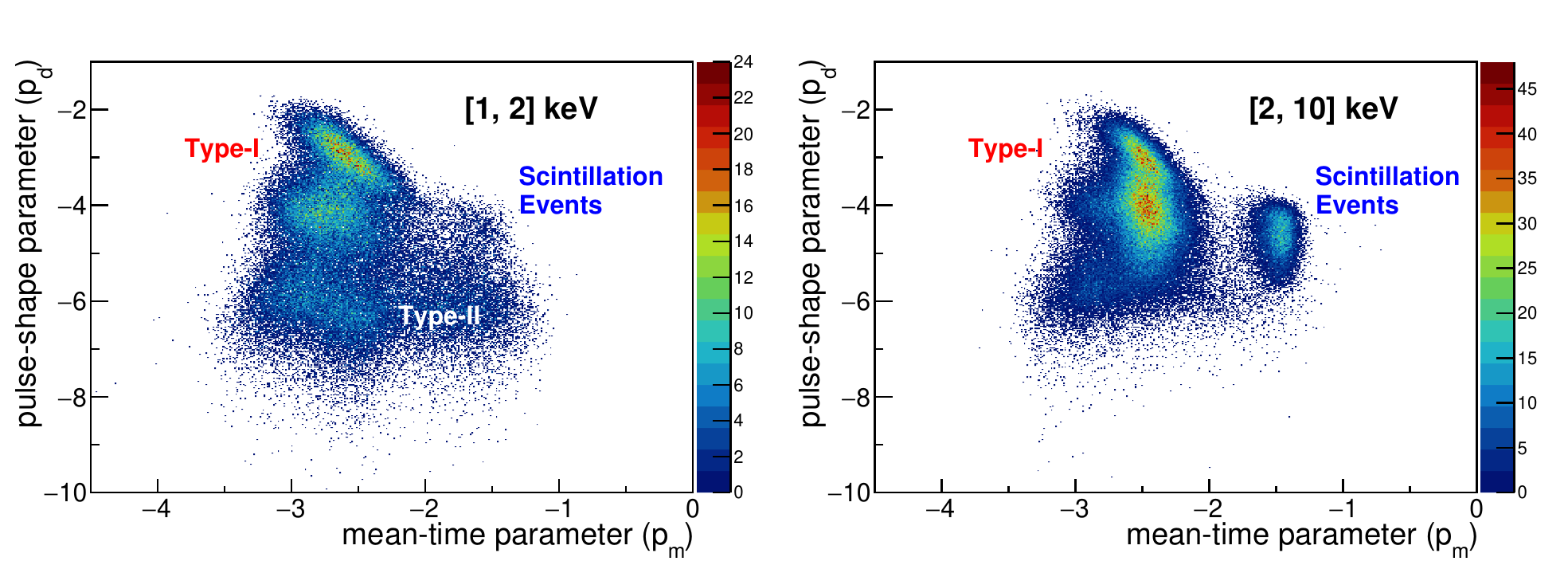}
\caption{Pulse-shape vs. mean-time parameter distributions for two different energy regions from the physics search data (59.5 days). The left (right) panel shows the distribution with the range of 1-2~(2-10)~keV. The new type of noise events does not appear above the 2~keV region as shown in the right plot and is the main source of noise events that are not separated from scintillation events at energies below 2~keV by only the mean-time parameter.}
\label{fig:dmPar}
\end{figure}

The new parameter that characterizes the PMT-pulse shape is defined as
\begin{eqnarray}
\lambda_{d,i} = -\frac{\ln\left(Q_{\mathrm{tail},i}/Q_{\mathrm{head},i}\right)}{T_{\mathrm{tail},i}-T_{\mathrm{head},i}},
\label{eq:SimplifiedDecayTime}
\end{eqnarray}
where $Q_\mathrm{head}\equiv \sum_{t_i < t_c} q_i$ and $Q_\mathrm{tail}\equiv \sum_{t_i > t_c} q_i$ are amplitude sums of the first and second half\footnote{The first and second half are divided with an event time span between the trigger time and the time of the last single photoelectron pulse within a 8~$\mathrm{\mu s}$ window.} in time of the $i^{th}$ PMT, respectively, and $T_\mathrm{head}\equiv (\sum_{t_i < t_c} q_i\cdot t_i)/Q_\mathrm{head}$ and $T_\mathrm{tail}\equiv (\sum_{t_i > t_c} q_i\cdot t_i)/Q_\mathrm{tail}$ are amplitude-weighted mean times in the first and second half, respectively. As shown in Fig.~\ref{fig:pulse_shape_parameter}, $q_i$ and $t_i$ in those definitions are charge and time of the $i^{th}$ photoelectron signal, and $t_c\equiv(t_1+t_n)/2$ is an average time of first and last photoelectron. The parameter $\lambda_d$ denotes a decay constant of two points ($T_{\mathrm{head},j}$,~$Q_{\mathrm{head},j}$) and ($T_{\mathrm{tail},j}$,~$Q_{\mathrm{tail},j}$) and quantifies the shape characteristic by focusing on the leading and trailing edge of a pulse. Again, the PMT-based $\lambda_d$ values are merged into a crystal parameter $p_d$, called the pulse-shape parameter, as $p_{d} \equiv \ln\left(\sum_i{\lambda_{d,i}}\right)$. Figure~\ref{fig:dmPar} shows event distributions by the mean-time parameter against the pulse-shape parameter for two different energy regions separately.

\subsection{PMT noise events in the 1--2 keV region}

While categorizing events using the two discrimination parameters above, we have found that a new population of noise events starts to appear at energies below 2~keV.  The leading-edge shape of the waveform from this new noise is the same as the previously identified PMT noise pulses, namely thin pulses, but the trailing part of the waveform is different and is more elevated from the baseline compared to the other pulses. We call this new noise a ``type-II PMT noise'' or a heavy-tail pulse. The exact origins of these PMT noises have not been fully understood at the moment.

By comparing the two plots in Fig.~\ref{fig:dmPar}, one can see that the categorized populations have little dependence in energy. In other words, the scintillation events and thin pulse PMT events stay in the same regions of the parameter space. On the other hand, the type-II noise events appear only at the right bottom corner in the 1--2 keV region.

\begin{figure}[t]
\centering
\includegraphics[width=0.8\textwidth]{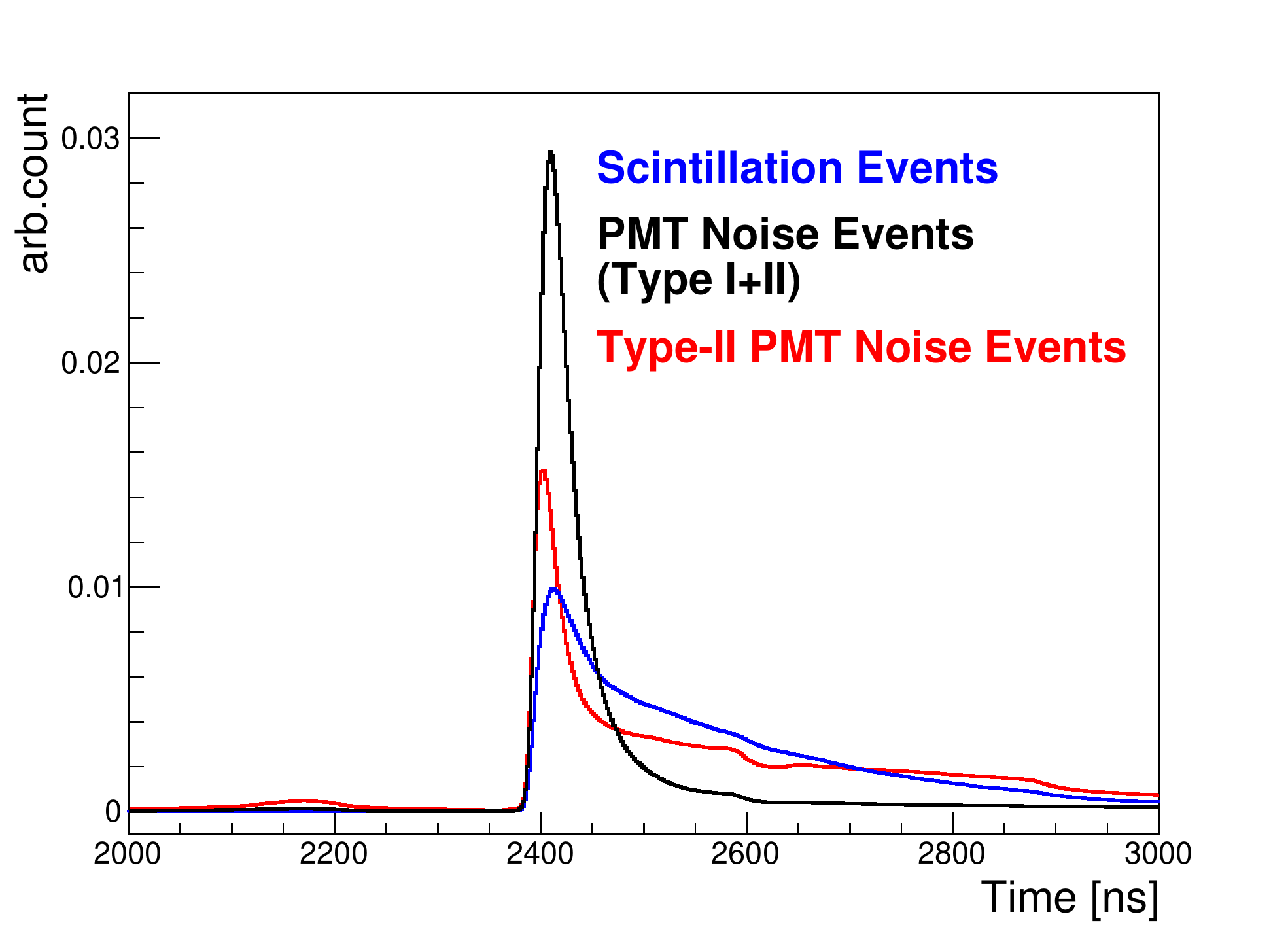}
\caption{Scintillation (blue) and noise event (black) shape templates. The templates are normalized probability density functions. The red waveform is the accumulated waveform of new type noise events extracted from PMT-noise events}
\label{fig:accWave}
\end{figure}

Both type-I and type-II noises contain the characteristic sharply-peaked structure at the leading edge time while the heavy-tail pulse includes a much slower tail compared to that of the type-I noise pulse as shown in Fig.~\ref{fig:accWave}. This type of events was not easily identified in the mean-time parameter because their elevated tails skewed the amplitude-weighted mean time towards higher values mimicking the scintillation events. On the other hand, the pulse-shape parameter uses the leading and trailing part of the waveform shape separately making this new type of noise visible in the parameter space. Although these parameters are still mediocre in terms of their separation power, we recognize that the shape of the waveforms can be further exploited. Therefore, to increase the scintillation signal purity, we developed a better selection parameter which utilizes the full waveform information.

\subsection{Likelihood parameter}
\begin{figure}[t]
\centering
\includegraphics[width=0.9\textwidth]{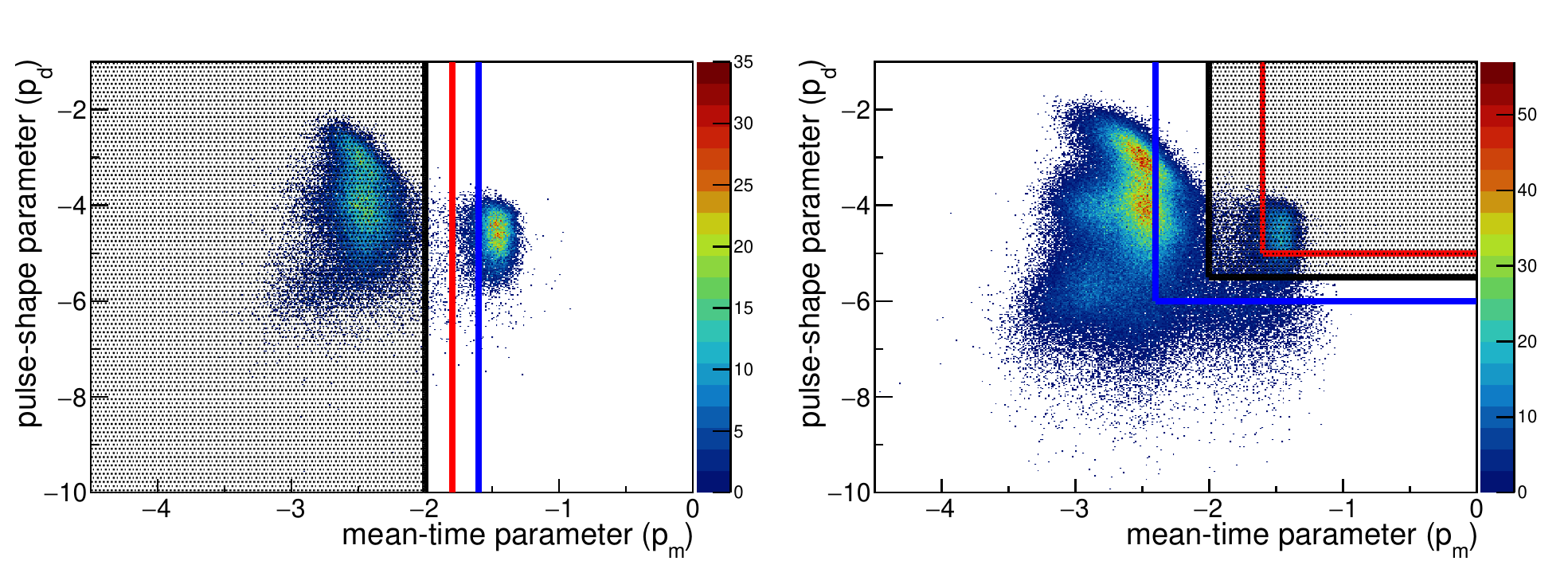}
\includegraphics[width=0.9\textwidth]{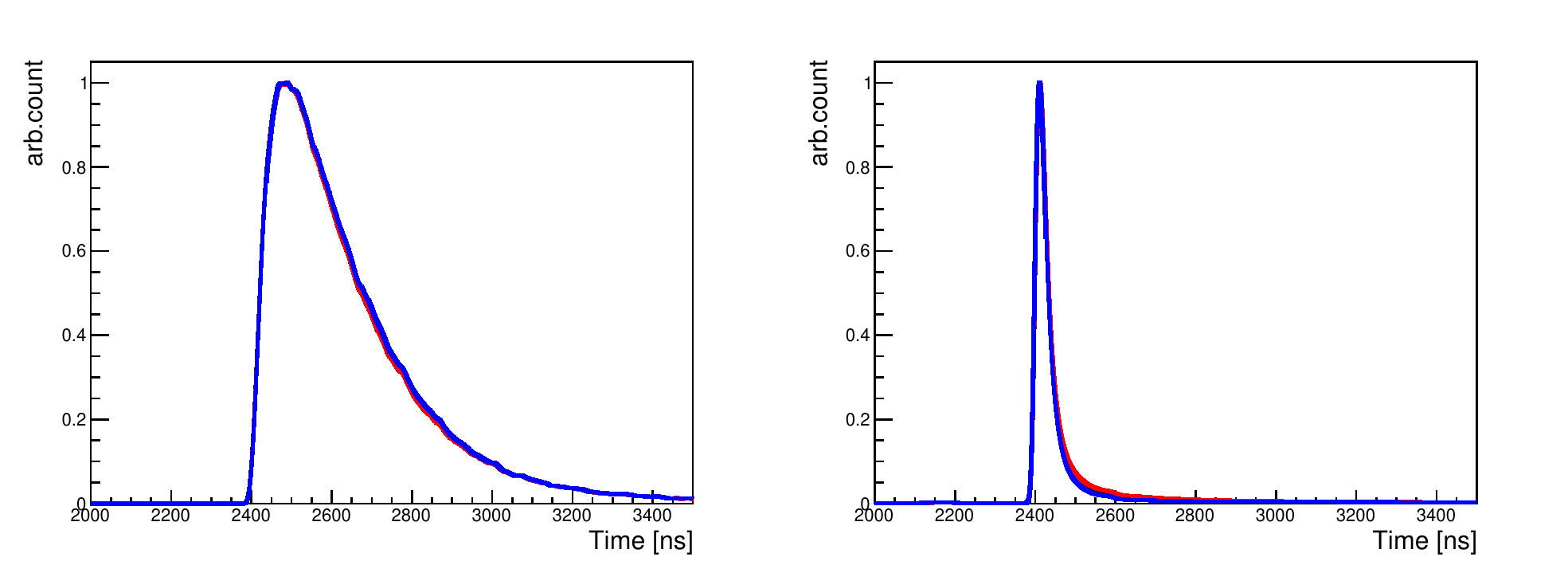}
\caption{(Top) Pulse-shape and mean-time parameter distributions for the $^{60}$Co-calibration data (left) and the physics-run data (right). Black lines in the top figures show the criteria to select events for reference waveforms of scintillation (left) and noise (right) event shape templates. The red and blue lines were chosen to validate the effect that variation of criteria might have on the reference waveforms. In the left figure, the black, red and blue lines are $p_m > -2$, $p_m > -1.8$ and $p_m > -1.6$, respectively, while the same color lines in the right figure are ($p_m > -2$ or $p_d < -5.5$), ($p_m > -1.6$ or $p_d < -5$) and ($p_m > -2.4$ or $p_d < -6$). The bottom figures show the shapes of the selected events according to each criterion. The color used is the same as the top figures.}
\label{fig:criteria_for_reference}
\end{figure}

Here, we construct a likelihood parameter that characterizes the PMT pulses using a template matching method. The definition of $\lambda_d$ in Eq.~\ref{eq:SimplifiedDecayTime} requires that the PMT pulse is divided into first (head) and second (tail) halves. Thus, each of the crystal's two PMT pulses has to contain two or more single-photoelectron hits, and this becomes an efficiency issue especially for the low-energy events below 2~keV where the number of measured photoelectrons is relatively low. In addition, the information contained in the waveform is partially exploited by the mean-time and pulse-shape parameters and so a single metric that computes likelihoods using a full waveform matching with a signal template and a noise template is preferable.

In order to obtain Compton-scattered low-energy events as a pure scintillation signal sample, data were taken for 27.9~days using a $^{60}$Co calibration source. Here, a noise-free sample of $e/\gamma$-induced scintillation signals can be extracted from multiple-hit events, defined as coincident-hit events with more than two crystals. We select events with a mean-time parameter cut ($p_m > -2$) only as shown in the top-left plot in Fig.~\ref{fig:criteria_for_reference} and make scintillation-event reference waveforms from 5000 of those events. In order to construct the corresponding noise reference waveforms, all types of PMT-noise events are selected via criteria based on both parameters ($p_m < -2$ or $p_d < -5.5$), from the events in the physics-run data as shown in the top-right plot of the Fig.~\ref{fig:criteria_for_reference}.

To begin with a parameter construction, a logarithmic likelihood of a waveform summed over the two PMT pulses associated with each event is evaluated for the signal and noise reference waveforms using
\begin{eqnarray}
\ln\mathcal{L} = 
\sum_i\left[T_i-W_i+W_i\ln\frac{W_i}{T_i}\right],
\end{eqnarray}
where $T_i$ and $W_i$ are the summed heights of the $i^\mathrm{th}$ time bin in the waveform for the template and event, respectively. As shown in the bottom plots of Fig.~\ref{fig:criteria_for_reference}, the shapes of template waveforms are sufficiently stable\footnote{When the peak height is normalized to 1 as shown in bottom plots of Fig.~\ref{fig:criteria_for_reference} and compared through the sum of differences between the waveforms, the difference between the waveforms of scintillation events is less than 2\% of the difference between the waveforms of scintillation and noise events. In the case of noise templates, the difference is about 1.2\% of the difference between waveforms of scintillation and noise events.} to significant variations of the selection criteria and therefore, the log-likelihood also has little dependence on the specific choice of cuts.

\begin{figure}[t] 
\centering
\includegraphics[width=0.9\textwidth]{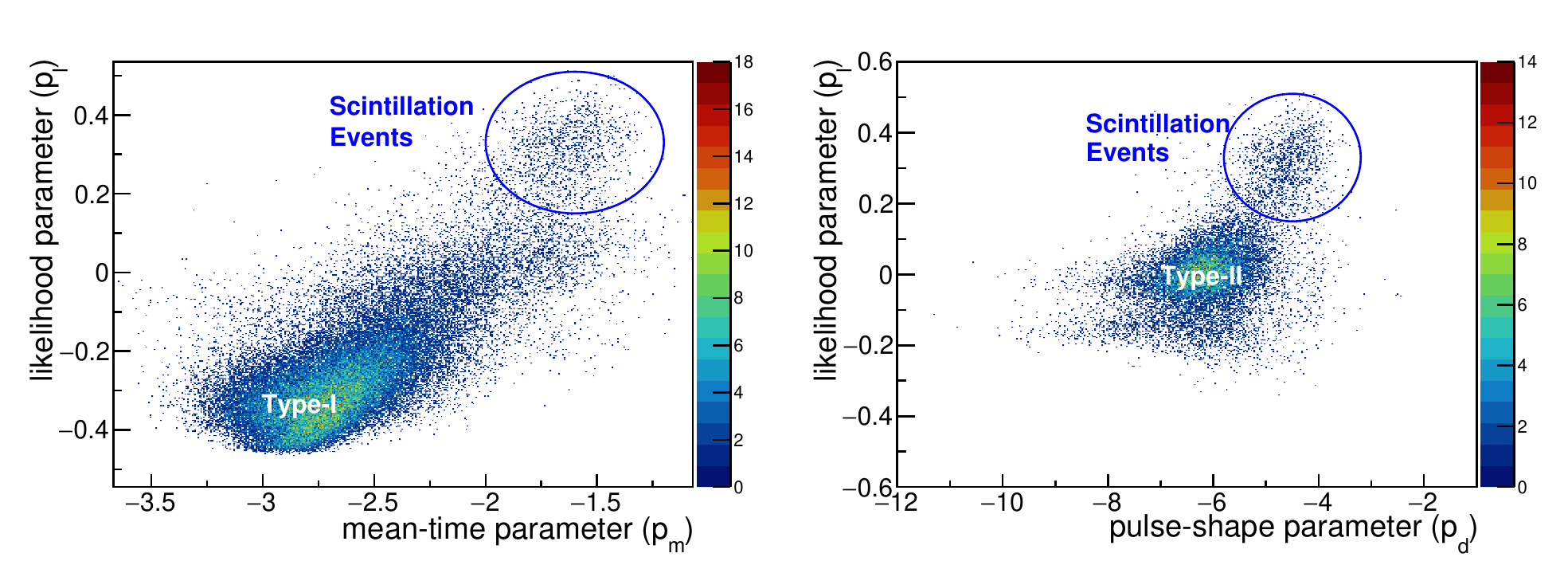}
\caption{(Left) Likelihood parameter and mean-time parameter distribution with the range of 1 to 2~keV and $p_d > -5.5$. (Right) Pulse-shape parameter and likelihood parameter distribution with the range of 1 to 2~keV and $p_m > -2$. Points are 59.5-days of physics-search data.}
\label{fig:pl_pm_pd}
\end{figure}

\begin{figure}[t] 
\centering
\includegraphics[width=0.8\textwidth]{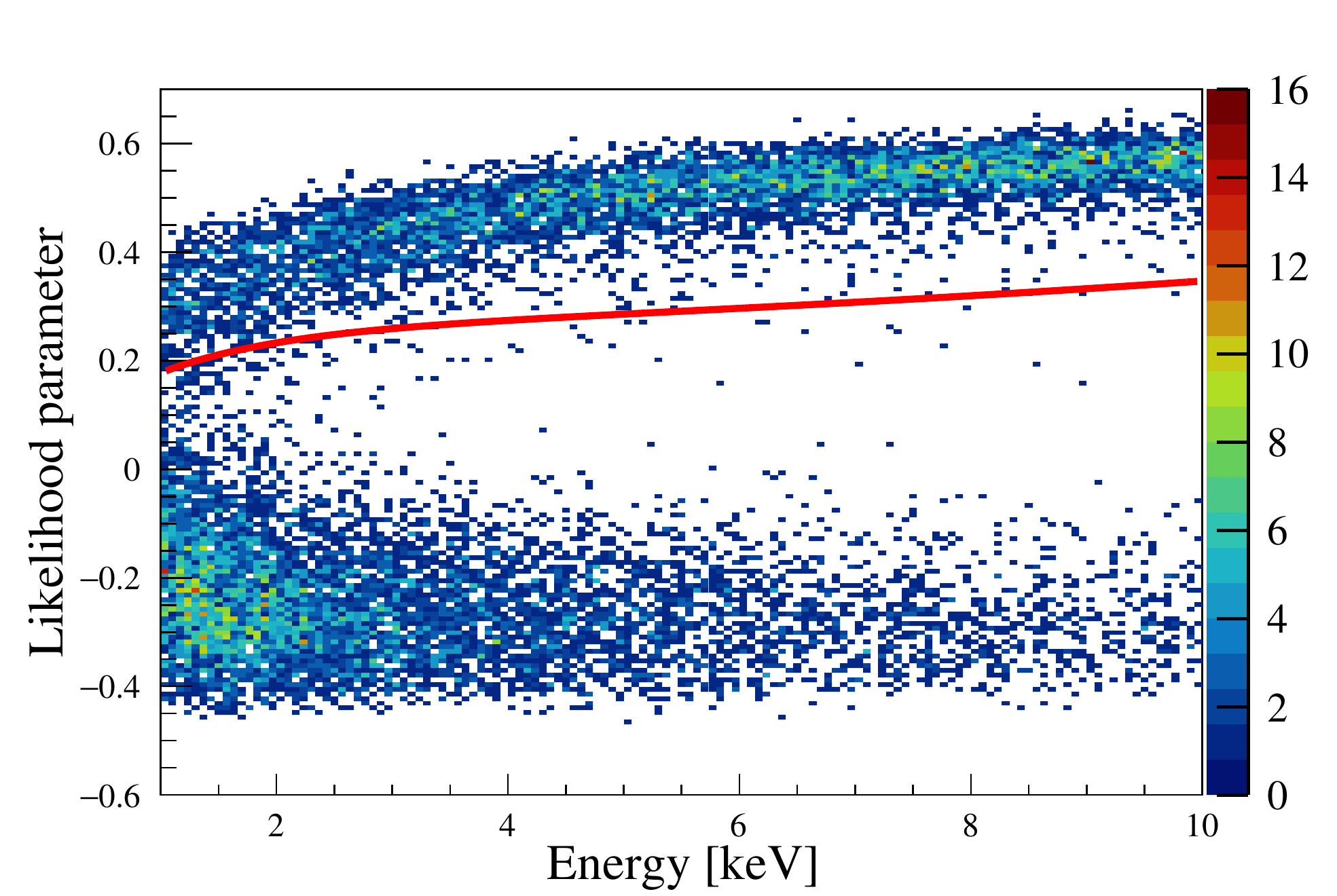}
\caption{Likelihood parameter ($p_l$) as a function of the energy for multiple-hit events in the $^{60}$Co-calibration data. The events above the red line are used to train the BDT.}
\label{fig:sigSample}
\end{figure}

We then have two logarithmic likelihood values for a crystal that are related to each of the two reference waveforms: scintillation and  the PMT-noise events. Next, we define a score as
\begin{eqnarray}
p_l = \frac{\ln\mathcal{L}_n - \ln\mathcal{L}_s}{\ln\mathcal{L}_n + \ln\mathcal{L}_s},
\label{eq:lpar}
\end{eqnarray}
where $\ln\mathcal{L}_s$ and $\ln\mathcal{L}_n$ denote logarithmic likelihoods obtained with scintillation-event and PMT-noise-event references, respectively. If an event has a small value of $\ln\mathcal{L}_s$ ($\ln\mathcal{L}_n$), it is more likely to be a scintillation (noise) event. Therefore, a large $p_l$ for an event implies that the event is more closely matched to the scintillation rather than the noise template. As shown in Fig.~\ref{fig:pl_pm_pd}, the likelihood parameter has a separation power that supersedes both the mean-time and pulse-shape parameters. In particular, as shown in the right plot of Fig.~\ref{fig:pl_pm_pd}, it has a stronger separation power for the type-II PMT noises than that of the pulse-shape parameter. The upper and lower bands in Fig.~\ref{fig:sigSample} denote the scintillation and the PMT-noise events, respectively, and demonstrate that this likelihood-based score parameter has separation capability in the 1 to 2~keV energy region.

\section{Machine learning technique for 1~keV threshold}
\label{sec:bdt}

For more efficient noise separation, we adopt a machine learning algorithm based on the parameters developed above. A Boosted Decision Tree (BDT) method that accounts for the correlations between individual parameters is efficient in combining several weak discriminating parameters into a single powerful discriminator. We trained a BDT to further reject the low energy PMT-noise events. The decision tree undergoes multiple iterations of trial selections based on the input variables associated with features of the scintillation events and PMT-noise events. As the iteration proceeds, based on the efficiency and purity of scintillation events in the previous event sample, the selections are improved and the BDT is trained on subsequent events with this importance applied. Eventually, a single discriminating parameter is created by combining the various selections according to their corresponding importance as a BDT score~\cite{BDT,hoecker2007tmva}. It should be noted that the BDT in this paper is updated relative to the BDT described in previous COSINE-100 publications~\cite{Adhikari:2018ljm,Kang:2019fvz} by the inclusion of additional discrimination parameters. The input BDT parameters used in the previous analysis are summarized in Ref.~\cite{Adhikari_2019} upon which we have updated two parameters by changing the $MT$ (Eq.~(3.6) in the reference) to mean-time parameter, $p_m$ and by adding the likelihood-based score parameter, $p_l$.

\subsection{Event selection for BDT training}

\begin{figure}[t]
\centering
\includegraphics[width=0.45\textwidth]{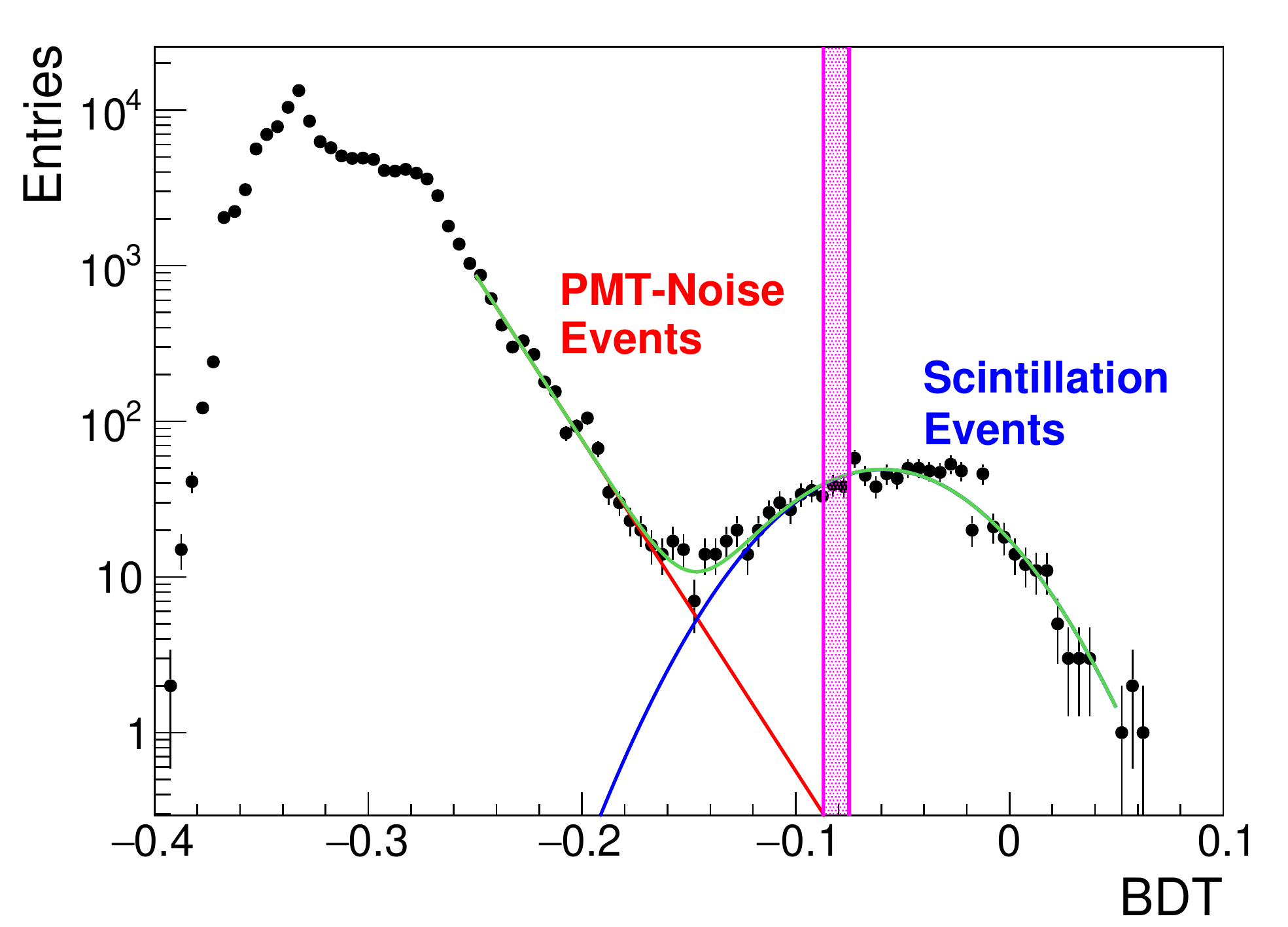}
\includegraphics[width=0.45\textwidth]{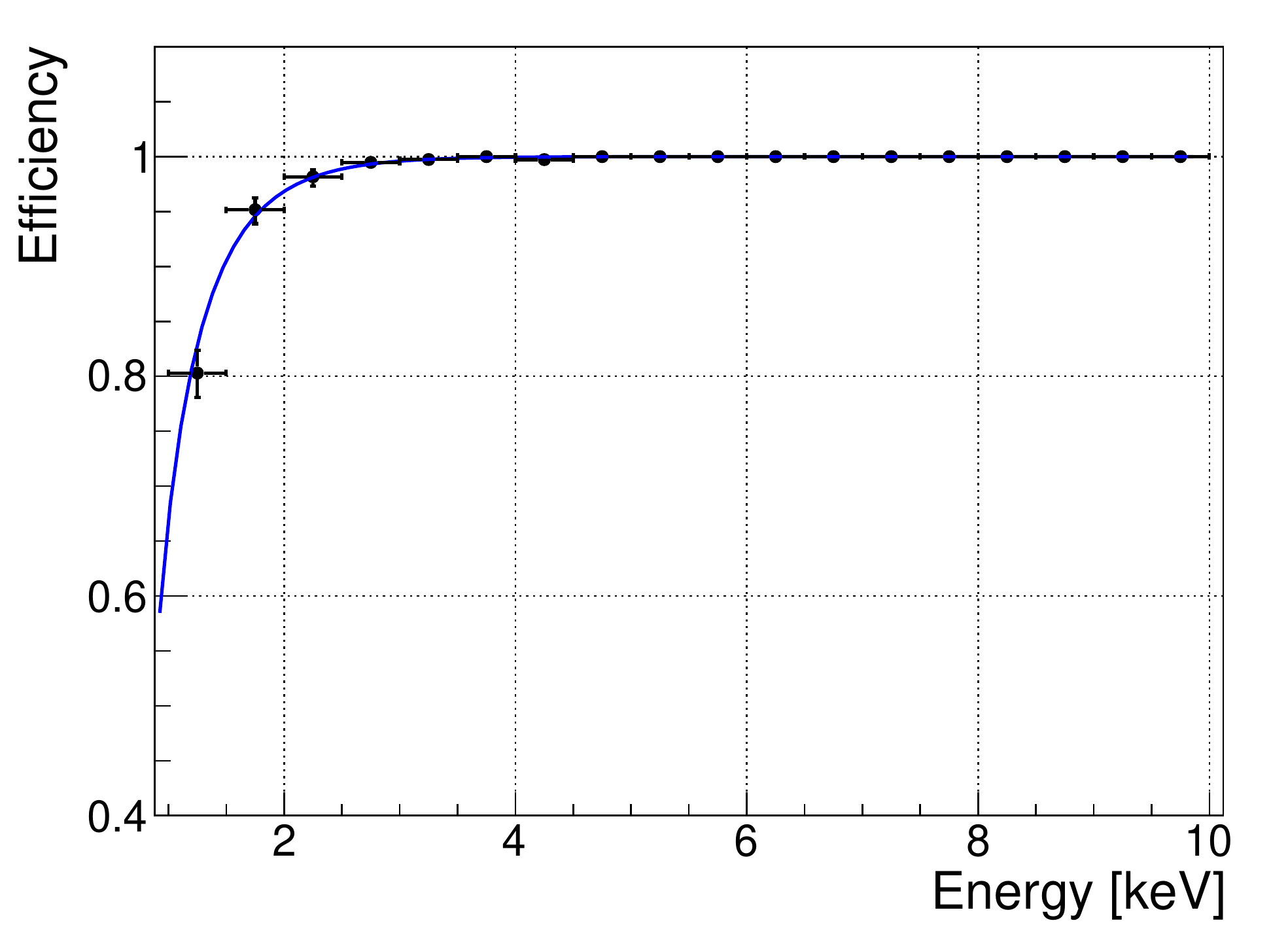}
\caption{The BDT distribution for events in the energy range of [1, 1.5]~keV (left) and the efficiency of scintillation events (right). In the left plot, the blue and red lines are fitted for scintillation events and PMT-noise events, respectively. The magenta region shows minimum/maximum lower limits of the red curve shown in Fig.~\ref{fig:bdtSelection}. In the case corresponding to the full line of Fig.~\ref{fig:bdtSelection}, the purity of scintillation events in the upper region of the criterion is more than 99.8\%. The black dots on the right plot shows the corresponding scintillation event efficiencies for the BDT criterion and the blue line is a fitted line with a cumulative beta function.}
\label{fig:bdtJustification}
\end{figure}

\begin{figure}[t]
\centering
\includegraphics[width=0.45\textwidth]{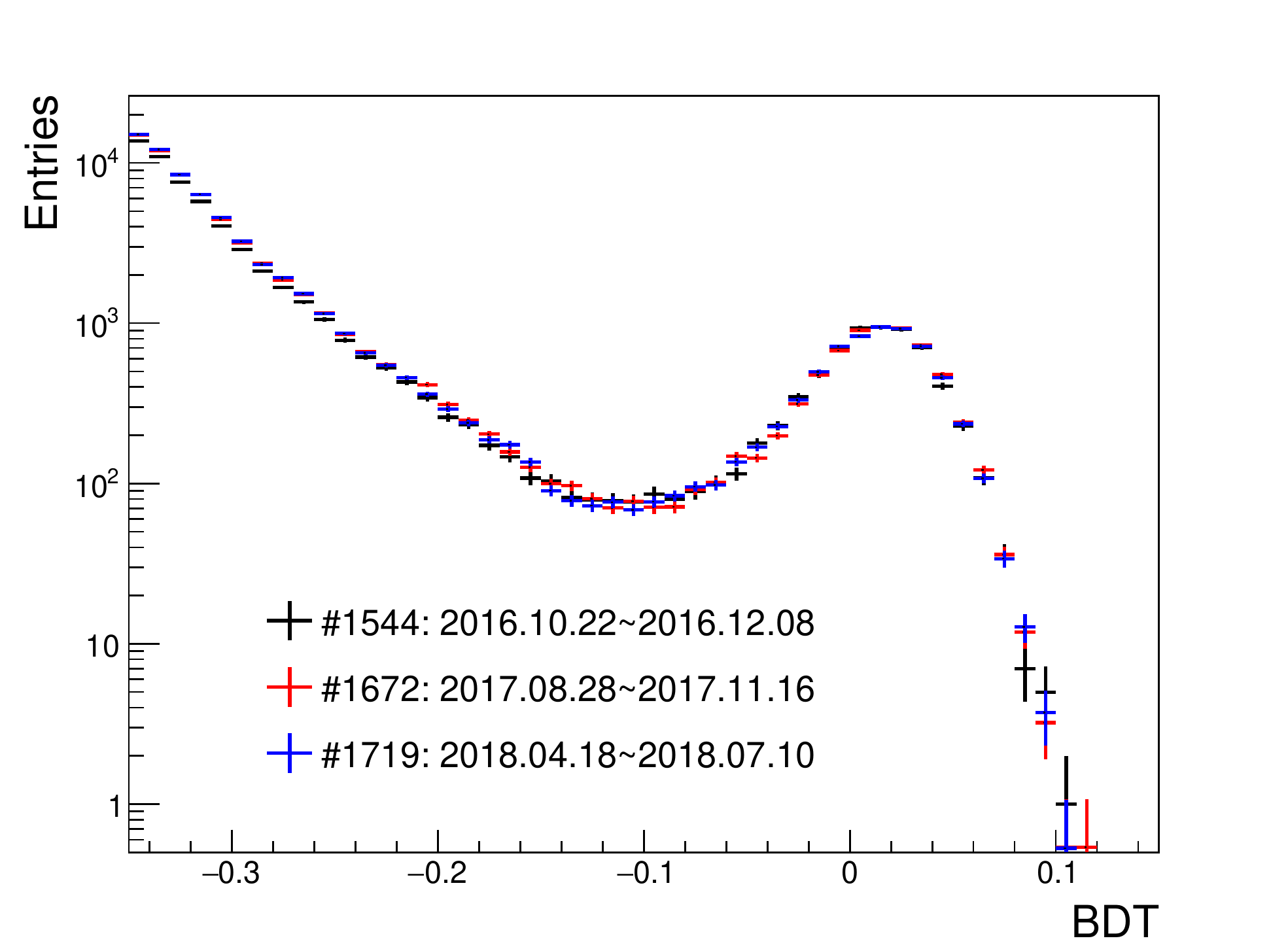}
\includegraphics[width=0.45\textwidth]{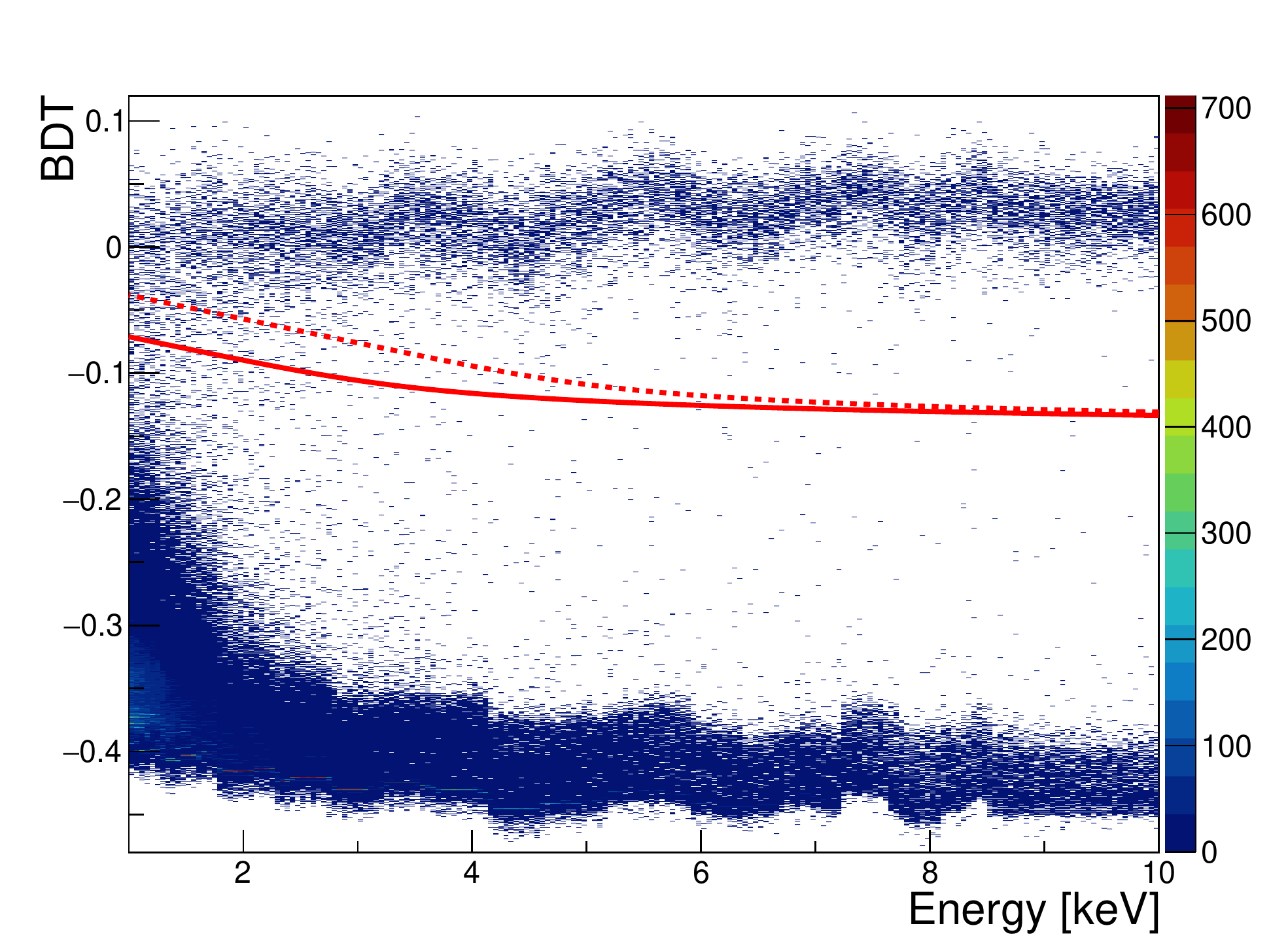}
\caption{BDT distribution of three data sets with similar time intervals (left) and BDT as a function of the energy in first 59.5 days of dark matter search data (right). The solid red line is the energy-dependent event selection where events above the red line are scintillation-like events that have less than 1\% noise contamination in the energy range from 1 to 1.5~keV and negligible noise contamination at higher energies. The dashed red line is a tighter cut to test the effect of the change in efficiency on sensitivity, as shown in Fig.\ref{fig:Sensitivity}}
\label{fig:bdtSelection}
\end{figure}

A challenging aspect of a BDT training is to obtain pure event samples that are used to model the scintillation and PMT-noise events. The Compton scattering events of $\gamma$-rays from a $^{60}$Co source (the events above the red line in Fig.~\ref{fig:sigSample}) are used as pure scintillation events in the BDT training. We estimate the scintillation event purity to be more than 99\% at the energy region between 1.0 and 1.5~keV by extrapolating the noise distribution (red Exponential fit) into the signal region (blue Gaussian line) as shown in Fig.~\ref{fig:bdtJustification}. For this energy region, we obtain a signal efficiency of about 80\% by evaluating the ratio of the number of events passed the cut to the total number of events using the calibration data. The first 59.5 days of the physics-run data, which are dominantly PMT noise-like events, are used as the noise sample for training the BDT, with 50\% of the initial data randomly sampled. The sample for the noise corresponds 5\% of the full analysis data and therefore little bias is expected. We find no time variation for our BDT selection as shown in the left plot of Fig.~\ref{fig:bdtSelection}. The BDT score as a function of energy of the physics-run data shown in the right plot of Fig.~\ref{fig:bdtSelection} exhibits a clear separation between scintillation and PMT-noise events for energies greater than 1~keV. The events above the red line are selected as scintillation-like events.

\subsection{Re-weighting the calibration variables for validation of the BDT}

\begin{figure}[t]
\centering
\includegraphics[width=0.8\textwidth]{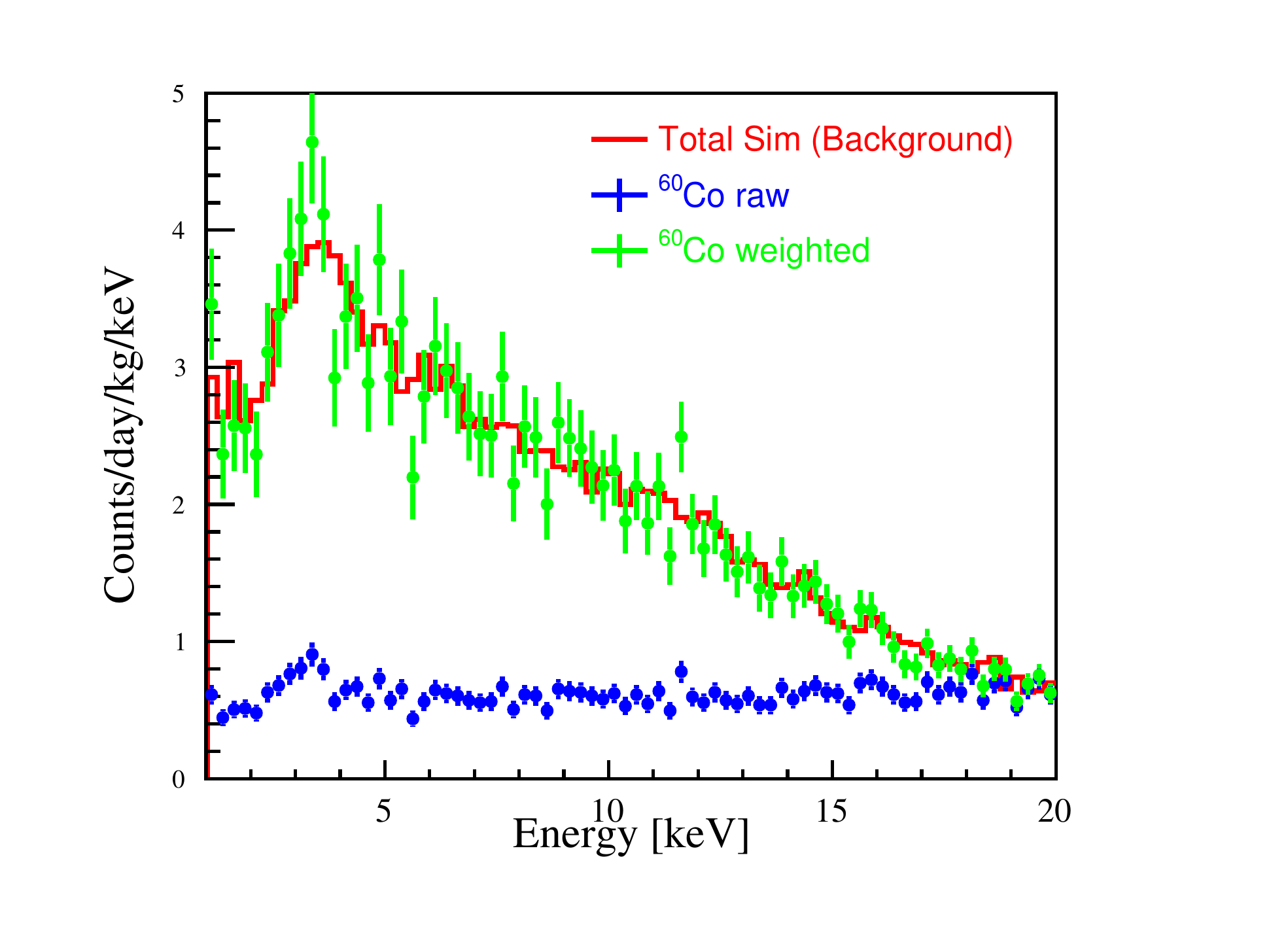}
\caption{Energy spectra for $^{60}$Co multiple-hit events before and after weighting. The blue data points indicate the raw $^{60}$Co spectrum while the green data points show the distribution with the Monte-Carlo energy (solid red line) weighting applied.}
\label{fig:reweight}
\end{figure}

Even with the good event separation, it is mandatory to validate the BDT and to quantify the selection efficiency. This would ensure that the events in the training calibration data behave the same way as those in the physics-run data. In order to validate the BDT training process, the input variables of the scintillation events in the $^{60}$Co-calibration data used for training the BDT are compared with those of the events selected from the independent physics-run data. As shown in Fig.~\ref{fig:reweight}, the energy spectrum of the $^{60}$Co-calibration data has a different shape from that for the physics-run data because the Compton scattered gammas are continuous in energy. Therefore, we apply Monte-Carlo calculated weights to the energy spectrum to match the background spectrum before making the comparison. The energy spectrum for the full simulation of the background radioisotopes is used to determine the spectrum weights. Figure~\ref{fig:reweight} shows the weighted spectrum from the $^{60}$Co-calibration data.

\begin{figure}[t]
\centering
\includegraphics[width=1.0\textwidth]{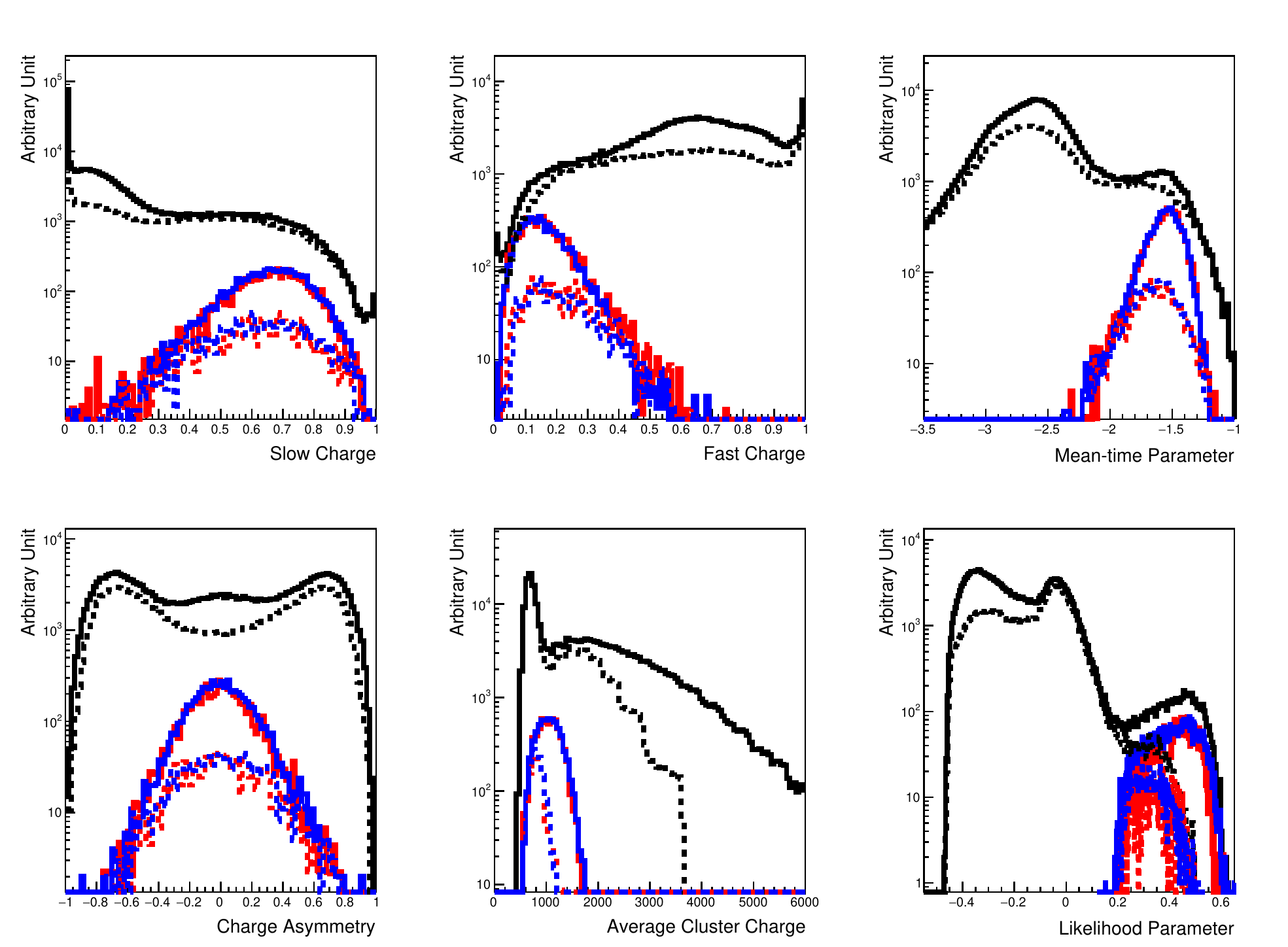}
\caption{Six example variables used to validate the BDT output response. The black, red and blue distributions are the total background, $^{60}$Co coincident events and scintillation-like events from the physics-run data, respectively. The solid and dotted lines denote [1,5] and [1,2] keV of energy ranges, respectively. All variables are energy-weighted.}
\label{fig:bdtValidation}
\end{figure}

Figure~\ref{fig:bdtValidation} shows the validation of the six input variables used to construct the BDT. The black line is the raw data while the blue line is the scintillation-like events selected by a BDT criterion from the physics-run data. The weights are applied to all selection variable distributions of the $^{60}$Co-calibration data to make them suitable for modeling the scintillation sample. After the BDT selection, there is good agreement in the variables between the weighted $^{60}$Co data and the selected scintillation data as shown in Fig.~\ref{fig:bdtValidation}. The consistency between the two independent samples provides an indirect validation of the procedure. In addition to the meantime and likelihood parameters, the other variables are defined as the ratio of the integrated charge between 500~ns and~600 ns to the integrated charge for the first 600~ns (Slow Charge), the ratio of the integrated charge between 0~ns and~50 ns to the integrated charge for the first 600~ns (Fast Charge), the balance of the deposited charge from each of the two PMTs (Charge Asymmetry), and, the average charge of clustered pulses (Average Cluster Charge). The selection efficiency for the energy bin between 1 and 1.5~keV is determined from the cut-applied $^{60}$Co sample divided by the original sample. The average efficiency of the COSINE-100 crystals is about 80\% and the efficiencies are distributed in the 70 to 88\% range.

\subsection{Sensitivity improvement with 1 keV threshold}
\label{sec:sensitivity}
In order to study the sensitivity of the annual modulation search with the 1~keV threshold, Monte Carlo experiments are used to calculate projected limits for the COSINE-100 detector in the case of no observed annual modulation signal. We assume a two years running time with a 3-counts/kg/day/keV flat background  (which excludes the two low-light-yield high-background crystals). The simulated data are fitted to a sinusoidal function with a fixed period and phase of one year. The fit is used to determine the simulated modulation amplitude observed by COSINE-100 at nuclear recoil energies ranging from 1-20 keV. We find that the DAMA/LIBRA modulation signal region with the lowered 1 keV threshold can be directly challenged by COSINE-100 data, as shown in Ref.~\cite{Thompson:2017yvq}.

\begin{figure}[t]
\centering
\includegraphics[width=0.8\textwidth]{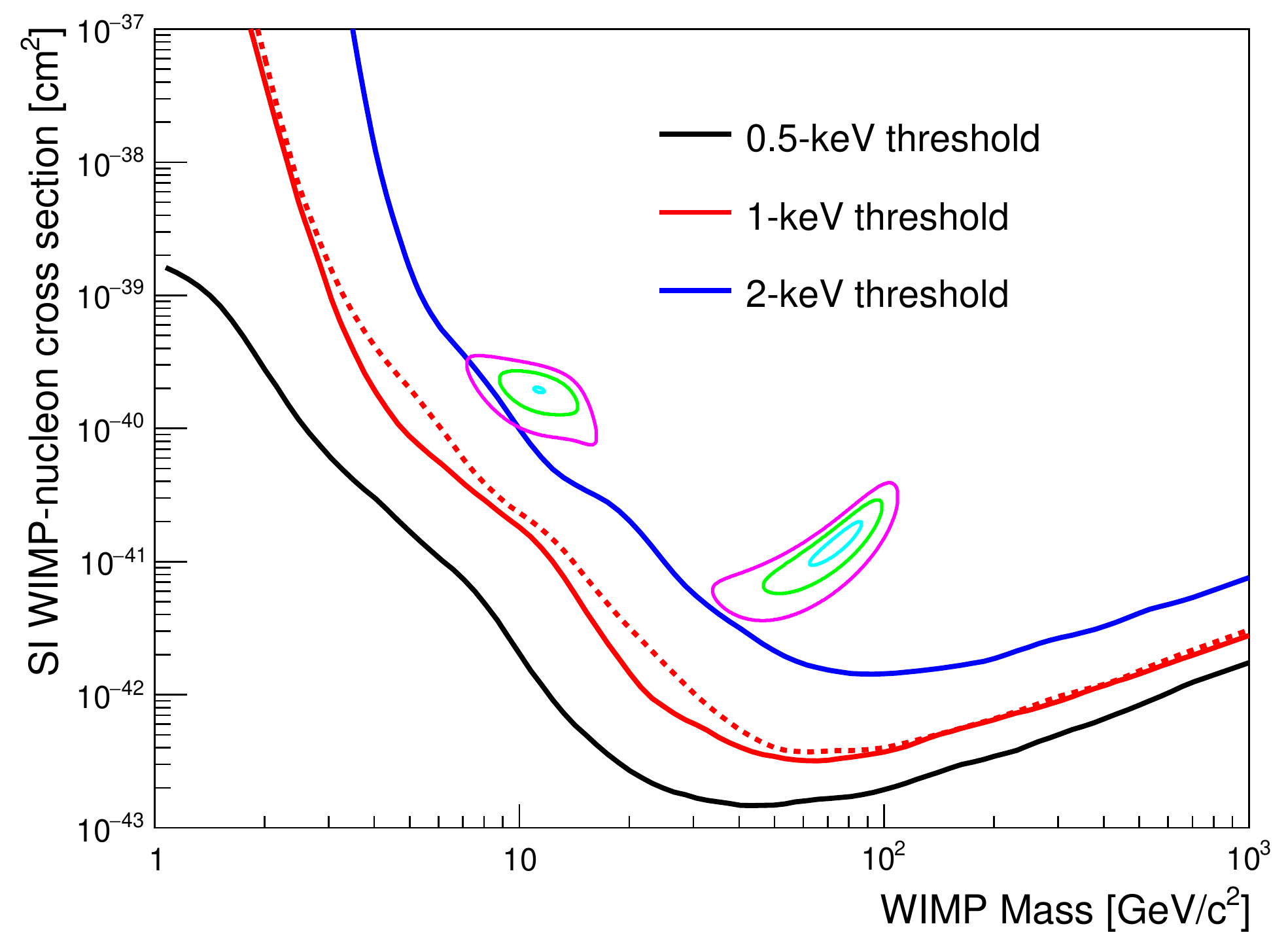}
\caption{The projected sensitivity of the COSINE-100 experiment with a 10000~kg$\cdot$day exposure, as derived from the constant rate observed above several thresholds. The black, red, and blue curves show the detector sensitivity for WIMP search with 0.5-, 1-, and 2-keV threshold, respectively. The dashed red line shows the sensitivity of 1-keV threshold with a tighter cut as shown in Fig.\ref{fig:bdtSelection}. The cyan, green, and magenta contours show 1, 3, and 5$\sigma$ regions, respectively, allowed by the DAMA/LIBRA-phase1 annual modulation signal.}
\label{fig:Sensitivity}
\end{figure}

Separately, the cross section parameter space for few-GeV/c$^2$ WIMP masses is relatively unexplored~\cite{Hehn_2016,PhysRevD.100.102002,PhysRevD.99.062001,PhysRevD.100.022001,PhysRevLett.121.081307} and there is growing interest in the low-mass WIMP search in the sub-GeV mass region. Therefore, the low energy event selection can provide improvement in the constant rate search using crystal detectors. In order to study WIMP-nucleon cross-section sensitivity as a function of the low mass WIMPs, we assume a 3-counts/kg/day/keV flat background that has 5\% overall systematic uncertainty. And the uncertainty from the efficiency estimation is also considered. A thousand pseudo-data sets based on the null hypothesis are used for the sensitivity estimation, where the assumed exposure is 10000~day$\cdot$kg. We also assume the isospin conserving spin-independent interaction and the halo model and conditions for generation of WIMP signal are the same as Ref.~\cite{Ko_2019}. Figure~\ref{fig:Sensitivity} shows COSINE-100 comparisons for different thresholds. The 1~keV threshold analysis shows a factor of ten improvement in sensitivity compared to the 2~keV threshold. To evaluate the stability and systematic impact of the selection indirectly, we show an additional sensitivity curve in Fig.~\ref{fig:Sensitivity} where we applied a tighter selection criterion presented in Fig.~\ref{fig:bdtSelection} as a dashed line. We find that the cut and the variation of the sensitivity are at most a factor two in some places. Additionally, we show a projected sensitivity for the low mass WIMPs with an assumed 0.5~keV threshold. Another factor of ten improvement for a 10~GeV/c$^2$ mass WIMP is expected compared with the 1~keV threshold analysis. To achieve this threshold, the development of additional procedures for the rejection of the remaining PMT-noise events is on-going.

\section{Summary \& Outlook}
\label{sec:summary}
A new PMT-related noise rejection algorithm based on a likelihood estimator and BDT training procedure is developed for the COSINE-100 dark matter experiment, which has been collecting data for more than three years at the Yangyang underground laboratory. The likelihood parameters calculated using categorized noise templates and the particle scintillation template helped to reject noise events down to energies of 1~keV and possibly lower. The current challenge for accessing events below 1~keV is largely due to the low number of photoelectrons produced and existence of sources of PMT-noise events. Further developments in software and hardware are necessary.

With the improved energy thresholds, and more than 3.5 years of running, COSINE-100 data can be directly compared to the DAMA/LIBRA annual modulation signal. Additionally, using constant rate analysis by averaging the WIMP search data for a given exposure, a spin-independent interaction sensitivity study of COSINE-100 shows that a significant improvement for the low mass WIMP search can be achieved. This study enables us to perform important searches in light of effective field theory operators and velocity dependent dark matter distributions where there are parameter spaces consistent with the DAMA/LIBRA modulation signals. By lowering the threshold below 1~keV, the role of background in the DAMA/LIBRA data can be further understood. For example, the rate of 0.85~keV Na-22 X-rays background line and of phosphorescence would tell us how much cosmogenic contamination DAMA/LIBRA might contain. More refined analyses with larger exposure are forthcoming.

Furthermore, the method to reject noise events in the NaI(Tl) crystal detector can be utilized in low-threshold NaI(Tl) experiments for coherent neutrino-nucleus scattering~\cite{Akimov1123}. The crystals become interesting in terms of neutrino-nucleon coherent elastic scattering if the threshold can be lowered to 0.5 keV with sufficient noise rejection. The same crystals can be used in the neutrino property measurement with high flux neutrinos, {\it{e.g.}} from a nuclear reactor or supernova.

\section*{Acknowledgments}
We thank the Korea Hydro and Nuclear Power (KHNP) Company for providing underground laboratory space at Yangyang. This work is supported by: the Institute for Basic Science (IBS) under project code IBS-R016-A1 and NRF-2016R1A2B3008343, Republic of Korea; NSF Grants No. PHY-1913742, DGE1122492, WIPAC, the Wisconsin Alumni Research Foundation, United States; STFC Grant ST/N000277/1 and ST/K001337/1, United Kingdom; Grant No. 2017/02952-0 FAPESP, CAPES Finance Code 001, CNPq 131152/2020-3, Brazil.

\bibliography{mybibfile}

\end{document}